\documentclass[3p]{elsarticle}
\pdfoutput=1
\usepackage[unicode=true, bookmarks=false,bookmarksnumbered=true, bookmarksopen=true, breaklinks=false, backref=false, pagebackref=false, colorlinks=true]{hyperref}
\hypersetup{pdftitle={Landau Gauge Fixing on GPUs}, pdfauthor={Nuno Cardoso, Paulo J. Silva, Pedro Bicudo and Orlando Oliveira}, linkcolor=black, citecolor=black, urlcolor=black, filecolor=black, pdfpagelayout=OneColumn, pdfnewwindow=true, pdfstartview=XYZ, plainpages=false}
\usepackage{color}
\usepackage{amsmath}
\usepackage{amssymb}
\usepackage{braket}
\usepackage[T1]{fontenc}
\usepackage{url}
\usepackage[english]{babel}
\usepackage{dcolumn}
\usepackage{bm}
\usepackage{subfig}
\usepackage{float}
\usepackage{url}
\usepackage{esint}
\usepackage{dsfont}
\usepackage{listings}
\usepackage{algorithm}
\usepackage{algorithmic}

\newcommand{\Tr}{\mbox{\rm Tr}}
\newcommand{\Real}{\mbox{\rm Re}}
\newcommand\T{\rule{0pt}{2.6ex}}
\newcommand\B{\rule[-1.2ex]{0pt}{0pt}}

\usepackage{verbatim}

\journal{Computer Physics Communications}

\begin{document}
\begin{frontmatter}

\title{Landau Gauge Fixing on GPUs}

\author[lis]{Nuno Cardoso\corref{cor1}}
\ead{nuno.cardoso@ist.utl.pt}

\author[coimb]{Paulo J. Silva}
\ead{psilva@teor.fis.uc.pt}

\author[lis]{Pedro Bicudo}
\ead{bicudo@ist.utl.pt}

\author[coimb]{Orlando Oliveira}
\ead{orlando@teor.fis.uc.pt}

\cortext[cor1]{Corresponding author}

\address[lis]{CFTP, Departamento de F\'{i}sica, Instituto Superior T\'{e}cnico, Universidade T\'{e}cnica de Lisboa, Av. Rovisco Pais, 1049-001 Lisbon, Portugal}

\address[coimb]{CFC, Departamento de F\'{\i}sica, Faculdade de Ci\^encias e Tecnologia, Universidade de Coimbra, 3004-516 Coimbra, Portugal}

%%%%%%%%%%%%%%%%%%%%%%%%%%%%%%%%%%%%%%%%%%%%%%%%%%%%%%%%%%%%%%%%%%%%%%%%%
%%%%%%%%%%%%%%%%%%%%%%%%%    abstract   %%%%%%%%%%%%%%%%%%%%%%%%%%%%%%%%%
%%%%%%%%%%%%%%%%%%%%%%%%%%%%%%%%%%%%%%%%%%%%%%%%%%%%%%%%%%%%%%%%%%%%%%%%%
\begin{abstract}
%PSILVA 
%In this paper we present and explore the performance of Landau gauge fixing in GPUs using CUDA.
%We implemented the steepest descent method algorithm with Fourier acceleration to circumvent the problem
%of critical slowing down in CUDA. The GPU code is compared to a MPI implementation running on a PC cluster.
%The CPU code is built using the Chroma library, a widely used library for the lattice calculations, 
%and parallel FFT routines from Michael Pippig. In what concerns the performance of the gauge fixing, 
%the computational power of a single Tesla C2070 GPU is equivalent to approximately 256 CPU cores.
In this paper we present and explore the performance of Landau gauge fixing 
in GPUs using CUDA. We consider the steepest descent algorithm 
with Fourier acceleration, and compare the GPU performance with a parallel 
CPU implementation.  Using $32^4$ lattice volumes, we find that the 
computational power of a single Tesla C2070 GPU is equivalent to 
approximately 256 CPU cores.  
\end{abstract}

\begin{keyword}
CUDA \sep GPU \sep Lattice Gauge Theory \sep SU(3) \sep Landau \sep Gauge Fixing
\MSC 12.38.Gc \sep 07.05.Bx \sep 12.38.Mh \sep 14.40.Pq
\end{keyword}

\end{frontmatter}

%%%%%%%%%%%%%%%%%%%%%%%%%%%%%%%%%%%%%%%%%%%%%%%%%%%%%%%%%%%%%%%%%%%%%%%%%
%%%%%%%%%%%%%%%%%%%%%%%%%    Introduction   %%%%%%%%%%%%%%%%%%%%%%%%%%%%%
%%%%%%%%%%%%%%%%%%%%%%%%%%%%%%%%%%%%%%%%%%%%%%%%%%%%%%%%%%%%%%%%%%%%%%%%%
\section{Introduction}

Quantum Chromodynamics (QCD) describes the interaction of quarks and gluons. The fundamental fields
of the theory have not been observed so far as free particles in nature. Quarks and gluons are confined inside 
hadrons, glueballs or other possible color singlet states. 
The description of the low energy properties of hadrons, such
%PSILVA h
as confinement and chiral symmetry breaking, clearly is beyond the reach of perturbative solutions of QCD.
The simulation of the theory on an hypercubic space-time lattice offers a first principle approach to solve
QCD.

The fundamental fields on the formulation of gauge theories on the lattice, the link variables,
live on the gauge group and not on the algebra. Further, the link variables are related to the 
usual continuum gauge fields via a non-linear transformation. For SU(N) gauge theories, the 
group is a compact manifold and, therefore, the lattice formulation does not require gauge fixing. 

Physical observables are gauge invariant and as long as one is interested only in gauge invariant operators, 
in principle, one can avoid gauge fixing. However, if one wants to understand the dynamics of the QCD 
fundamental fields of QCD via the computation of quark and gluon correlation functions, as well as
expectation values of operators between quark and gluon states, then one needs to choose a particular gauge
(Elitzur's theorem \cite{Elitzur:1975im}). For example, the non-perturbative computation of the renormalization constant of
composite operators can be realized by looking at the expectation values between quark states 
\cite{Martinelli1995}. The quark and gluon Green's functions allow for a first principles determination of the
QCD running coupling \cite{Oliveira2007,Bogolubsky2009}, a computation of the gluon running 
mass \cite{Oliveira2011}, of the quark running mass \cite{Bowman2005,Furui2006} or the non-perturbative
$\langle A^2 \rangle$ condensate \cite{Boucaud2009,Dudal2010}. 
Furthermore, gauge fixing has also been used as a way to smooth the gauge configurations, to improve the
signal to noise ratio and access various functions with smaller statistical errors.  For a general overview on 
gauge fixing in lattice gauge theories and related issues see \cite{Gusti2001}.

On the lattice, the most common procedure for lattice gauge fixing is defined by
%On the lattice, gauge fixing is defined by 
maximizing a functional of 
%PS the 
the link variables, generated by importance sampling,
 over the gauge orbits of each link. Recall that a gauge orbit is the set of all links obtained
from a given one by means of a gauge transformation. 
The gauge fixing problem requires finding the maxima of a functional 
%PS define
defined 
over a manifold with a very large
dimension. For sufficiently large lattices, the bottleneck of a simulation is actually the maximizing problem, 
i.e. the gauge fixing when required, rather than the importance sampling. Therefore, 
it is important to access fast algorithms which, for large 
%PS system, 
systems,
do not suffer from the problem of the critical
slowing down problems and profit from modern high performant computers including graphical processor units 
(GPU).

In the present work we discuss the implementation of Fourier accelerated steepest descent method for
Landau gauge fixing in lattice QCD \cite{Davies:1987vs} on GPUs using CUDA and compare
its performance with an MPI implementation based on the Chroma library.
%PSILVA The
This 
Landau gauge fixing algorithm has a dynamical critical exponent close to zero and should remain 
efficient for large lattices. In \cite{Cucchieri1998,Cucchieri1998b} strategies for parallelizing the algorithm
using MPI,  as well as possible alternative approaches to avoid relying on parallel fast Fourier transformations, 
were investigated. 

The paper is organized as follows. In section \ref{LatticeGF} we describe briefly the basics of lattice QCD
simulations together with the definitions required to gauge fix to the Landau gauge. In section \ref{paralelo}
%PS we detailed on 
we detail 
the parallel implementation of the Fourier accelerated steepest descent method using MPI
and CUDA. In section \ref{resultados} the GPU code is compared against an MPI implementation running
on a high performance cluster, 
%PS

using $32^4$ lattice volumes.
 Finally, in section \ref{conclusoes} we conclude.

%%%%%%%%%%%%%%%%%%%%%%%%%%%%%%%%%%%%%%%%%%%%%%
%%%%%%%%%%%%%%%%%%%%%%%%%    Lattice   %%%%%%%%%%%%%%%%
%%%%%%%%%%%%%%%%%%%%%%%%%%%%%%%%%%%%%%%%%%%%%%
\section{Lattice QCD and Landau Gauge Fixing \label{LatticeGF}}

In the lattice formulation of QCD, the fundamental fields are the link variables
\begin{equation}
  U_\mu (x) ~ = ~ \exp \left( i a g_0 A_\mu (x + a \hat{e}_\mu / 2) \right)
  \, ,
\end{equation}
where $\hat{e}_\mu$ are unit vectors along the $\mu$ direction and $A^a_\mu$ the usual gluon fields.
The links belong to the SU(3) group. QCD is a gauge theory and the fields related by gauge transformations
\begin{equation}
  U_\mu (x) ~ \longrightarrow ~ g(x) ~ U_\mu (x) ~ 
 g^\dagger (x + a \hat{e}_\mu) \, ,
 \hspace{1cm} g \in SU(3) \, ,
\end{equation}
are physically equivalent. The set of fields related by gauge transformations
defines a gauge orbit and it is enough to pick one field from each of the orbits to study gauge theories.
The identification of one field in each gauge orbit is called gauge fixing.

On a typical simulation, an ensemble of gauge configurations, where a configuration is understood as the set 
of links defined on each space-time point, is generated via importance sampling using, for
example, the Wilson action. The gauge fixing is performed \textit{a posteriori} on each of the gauge configurations.
The Landau gauge is defined by maximizing the functional
\begin{equation}
	F_U[g]=\frac{1}{4 N_cV}\sum_x\sum_\mu\Real\left[ \Tr\left(  g(x)U_\mu(x)g^\dagger(x+\mu) \right) \right] \, ,
\end{equation}
where $N_c$ the dimension of the gauge group and $V$ the lattice volume, on each gauge orbit.
One can prove, \cite{Oliveira:2003wa}, that picking a maxima of $F_U[g]$ on a gauge orbit is equivalent to demanding the usual 
continuum Landau gauge condition $\partial_\mu A^a_\mu = 0$ and to require the positiveness of the Faddeev-Popov determinant. In the literature this gauge is known as the minimal Landau gauge.

The functional $F_U[g]$ can be 
%PS maximise
maximized using the steepest descent method
 \cite{Davies:1987vs,Oliveira:2003wa}. However, when the method is applied to large lattices, it
 faces the problem of critical slowing down, which can be attenuated by Fourier acceleration. 
 In the Fourier accelerated method, at each iteration one chooses
\begin{equation}
	g(x) = \exp\left[ \hat{F}^{-1} \frac{\alpha}{2}\frac{p^2_\text{max}a^2}{p^2a^2} \hat{F} \left( \sum_\nu \Delta_{-\nu}\left[U_\nu(x)-U_\nu^\dagger(x)\right]   -\text{trace} \right)\right]
	\label{eq:gx}
\end{equation}
with
\begin{equation}
	\Delta_{-\nu}\left(U_\mu(x)\right) = U_\mu(x-a\hat{\nu})-U_\mu(x)
\end{equation}
$p^2$ are the eigenvalues of $\left(-\partial^2\right)$, $a$ is the lattice spacing and $\hat{F}$ represents a 
fast Fourier transform (FFT).  For the parameter $\alpha$, we use the recommended  value 
0.08 \cite{Davies:1987vs}. In what concerns the computation of $g(x)$, for numerical purposes, it is enough 
to expand the exponential in equation (\ref{eq:gx}) to first order in $\alpha$, followed by a reunitarization, i.e.
a projection to the SU(3) group.

The evolution and convergence of the gauge fixing process can be monitored by
\begin{equation}
	\theta = \frac{1}{N_c V}\sum_x \Tr\left[\Delta(x)\Delta^\dagger(x)\right]
\end{equation}
where
\begin{equation}
	\Delta(x) = \sum_\nu\left[ U_\nu(x-a\hat{\nu})-U_\nu(x) - \text{h.c.} - \text{trace} \right]
\end{equation}
is the lattice version of $\partial_\mu A_\mu=0$ and $\theta$ is the mean value of
$\partial_\mu A_\mu=0$ evaluated over all space-time lattice points per color degree of freedom.
In all the results shown below, gauge fixing was stopped only when $\theta \le 10^{-15}$.

The Landau gauge fixing algorithm using FFTs is described in Algo. \ref{alg:fft}.

\begin{algorithm}[!h]
\caption{Landau gauge fixing using FFTs.}
\label{alg:fft}
\begin{algorithmic}[1] 
\STATE calculate $\Delta(x)$, $F_g[U]$ and $\theta$
\WHILE{$\theta \geq \epsilon$}
\FORALL{ element of $\Delta(x)$ matrix }
%\STATE apply FFT
\STATE apply FFT forward
\STATE apply $p^2_\text{max}/p^2$
%\STATE apply IFFT
\STATE apply FFT backward
\STATE normalize 
\ENDFOR
\FORALL{$x$}
\STATE obtain $g(x)$ and reunitarize
\ENDFOR
\FORALL{$x$}
\FORALL{$\mu$}
\STATE $U_\mu(x) \rightarrow g(x)U_\mu(x)g^\dagger(x+\hat{\mu})$
\ENDFOR
\ENDFOR
\STATE calculate $\Delta(x)$, $F_g[U]$ and $\theta$
\ENDWHILE
\end{algorithmic}
\end{algorithm}

%%%%%%%%%%%%%%%%%%%%%%%%%%%%%%%%%%%%%%%%%%%%%%%%%
%%%%%%%%%%%%%%%%%%%%%%%%%    Implementation   %%%%%%%%%%%%%%
%%%%%%%%%%%%%%%%%%%%%%%%%%%%%%%%%%%%%%%%%%%%%%%%%
\section{Parallel Implementation of the Gauge Fixing Algorithm \label{paralelo}}

%%%%%%%%%%%%%%%%%%%%%%%%%%%%%%%%%%%%%%%%%%%%%%%%%
%%%%%%%%%%%%%%%%%%%%%%%%%%%%%%%%%%%%%%%%%%%%%%%%%
%\subsection{MPI Implementation}
\subsection{CPU Implementation}

The MPI parallel version of the algorithm was implemented in C++, using the machinery provided by the Chroma library \cite{Edwards2005}. The Chroma library is built on top of QDP++, a library which provides a data-parallel programming environment suitable for Lattice QCD. The use of QDP++ allows the same piece of code to run both in serial and parallel modes. For the Fourier transforms, the code uses PFFT, a parallel FFT library written by Michael Pippig \cite{Pippig2011}. Note that, in order to optimize the interface with the PFFT routines, we have compiled QDP++ and Chroma using a lexicographic layout. 

%%%%%%%%%%%%%%%%%%%%%%%%%%%%%%%%%%%%%%%%%%%%%%%%%
%%%%%%%%%%%%%%%%%%%%%%%%%%%%%%%%%%%%%%%%%%%%%%%%%
%\subsection{CUDA Implementation}
\subsection{GPU Implementation}

For the parallel implementation of the SU(3) Landau gauge fixing pure gauge on GPUs we used
version 4.1 of CUDA. 

CUDA programs call parallel kernels, each of which executes in parallel across a set of
parallel threads. These threads are then organized, by the compiler or the programmer,
in thread blocks and grids of thread blocks.
The GPU instantiates a kernel program on a grid of parallel thread blocks. Within
the thread blocks, an instance of the kernel will be executed by each thread, which
has a thread ID within its thread block, program counter, registers, per-thread private
memory, inputs, and output results. Thread blocks are sets of concurrently executing
threads, cooperating among themselves by barrier synchronization and shared memory.
Thread blocks also have block IDs within their grids. A grid is an array of thread blocks.
This array executes the same kernel, reads inputs from global memory, writes results to
global memory, and synchronizes between dependent kernel calls.

For the 4D dimensional lattice, we address one thread per lattice site.
Although CUDA supports up to 3D thread blocks \cite{NVIDIA:cudac}, the same does not happen for the grid, 
which can be up to 2D or 3D depending on the architecture and CUDA version.
For grids up to 3D, this support happens only for CUDA version 4.x and for a CUDA device compute
capability bigger than 1.3, i.e. at this moment only for the Fermi and Kepler architectures.
Nevertheless, the code is implemented with 3D thread blocks and for the grid we adapted 
the code for each situation. 
Since our problem needs four indexes, using 3D thread blocks (one for t, one for z and one for both x and
y), we only need to reconstruct the other lattice index inside the kernel.
We use the GPU constant memory to put most of the constants needed by the GPU, like the number of points in the lattice, using \lstinline!cudaMemcpyToSymbol()!.

To store the lattice array in global memory, we use a SOA type array as described in \cite{Cardoso:2011xu}. 
The main reason to do this is due to the FFT implementation algorithm, Algo. (\ref{alg:fft}). 
The FFT is applied for all elements of the $\Delta(x)$ matrix separately. Using the SOA type array, the FFT can be applied directly to the elements without the necessity of copying data or data reordering.

In order to apply the Fast Fourier Transform in CUDA, we use CUFFT \cite{CUFFT}, 
the NVIDIA CUDA Fast Fourier Transform (FFT) library. The FFT is a divide-and-conquer algorithm for efficiently
computing discrete Fourier transforms of complex or real-valued data sets. 
The CUFFT library provides a simple interface for computing parallel FFTs on an NVIDIA GPU, which allows users 
to leverage the floating-point power and parallelism of the GPU without having to develop a custom, CUDA FFT 
implementation.

While the CUFFT library by Nvidia supports 1D, 2D and 3D FFTs, there is no direct support for 4D FFTs. 
As the FFT is cartesian separable, it is however possible to do a 4D FFT by applying four consecutive 1D FFTs. 
CUFFT has currently four different types of FFT plan configuration, \lstinline!cufftPlan1d()!, \lstinline!cufftPlan2d()! 
and \lstinline!cufftPlan3d()! for 1D, 2D and 3D respectively. Finally the \lstinline!cufftPlanMany()! can be used
in 1D, 2D and 3D with support for advanced data layouts.
While \lstinline!cufftPlan1d()! and \lstinline!cufftPlanMany()! allow the execution of multiple independent FFTs in parallel, in \lstinline!cufftPlan2d()! and \lstinline!cufftPlan3d()! this is not possible.

To perform a 4D FFT, a batch of 1D FFTs are applied along the first dimension in which the data is stored;
i.e., along x if the data is stored as (x, y, z, t).
 Between each 1D FFT, it is thereby necessary to change the order of the data;
 e.g., from (x, y, z, t) to (y, z, t, x). 
The drawback with this approach is that the time it takes to change the order of the data can be longer than to 
actually perform the 1D FFT. 
Using 3D FFT plus 1D FFT or two 2D FFTs, we can avoid the necessity to change the order of the data several 
times compared with using only 1D FFTs.
Therefore, with \lstinline!cufftPlanMany()!, which supports launching a batch of 1D, 2D and 3D FFTs, it is sufficient 
to change the order of the data once, and the same to perform the inverse FFT (IFFT).

The Landau gauge fixing using FFTs is described in Algo. (\ref{alg:fft}).
The Landau gauge fixing implementation with FFTs  requires more memory than the overrelaxation 
algorithm, see, for example, \cite{Schrock:2011hq}. 
Indeed, in addition to the lattice gauge array, it is necessary to have 
three more arrays to store $\Delta(x)$, $g(x)$ (with one fourth of the lattice array size) and to perform a data 
ordering (a complex array with the size of the lattice volume).

In order to reduce memory traffic we can use the unitarity of SU(3) matrices and store only the first two rows (twelve real numbers) and reconstruct the third row on the fly when needed, instead of storing it,
\begin{equation}
\begin{array}{cccccc}
\begin{pmatrix}a_1 & a_2 & a_3\\
b_1 & b_2 & b_3\\
c_1 & c_2 & c_3
\end{pmatrix} & \in \text{SU(3)}, & & \mathbf{c}=(\mathbf{a}\times \mathbf{b})^*\end{array}
\end{equation}
Although, in SU(3), we can use eight real numbers, this approach proves to be numerically unstable, \cite{Clark:2009wm,Babich:2010mu}.

To perform Landau gauge fixing in GPUs using CUDA, we developed the following kernels:
\begin{itemize}
\renewcommand{\labelitemi}{$\bullet$}
	\item k1: kernel to obtain an array with $p^2_\text{max}/p^2$.
	\item k2: kernel to calculate $\Delta(x)$, $F_g[U]$ and $\theta$. The sum of $F_g[U]$ and $\theta$ over all the lattice sites are done with the parallel reduction code in the NVIDIA GPU Computing SDK package \cite{parallelreduction}, which is already an optimized code.
	\item k3: kernel to perform a data ordering.
	\item k4: apply $p^2_\text{max}/p^2$ and normalize.
	\item k5: obtain $g(x)$ and reunitarize.
	\item k6: perform  $U_\mu(x) \rightarrow g(x)U_\mu(x)g^\dagger(x+\hat{\mu})$.
\end{itemize}
In Table \ref{tab:memflop}, we show the number of floating-point operations and the number of memory loads/stores per thread by kernel. The number of floating-point operations using 2D plus 2D FFTs is given by $nx\times	ny\times nz\times nt\times 5( \log_2(nx\times ny)+\log_2(nz\times nt))$.

\begin{table}[!h]
\begin{center}
\begin{tabular}{|c|c|c|c|c|}
\hline
\T\B \textbf{kernel}	&	\multicolumn{2}{c|}{18real}	&	\multicolumn{2}{c|}{12real}	\\ \hline
\T\B per thread	&	load/store	& flop	&	load/store	& flop	\\ \hline
\hline
\T\B k1	&	0/1	&	20	&	0/1	&	20	\\ \hline
\T\B k2	&	144/20	&	505	&	96/14	&	841	\\ \hline
\T\B k3	&	2/2	&	0	&	2/2	&	0	\\ \hline
\T\B k4	&	3/1	&	2	&	3/1	&	2	\\ \hline
\T\B k5	&	18/18	&	153	&	12/12	&	153	\\ \hline
\T\B k6	&	162/72	&	1584	&	108/48	&	1962	\\ \hline
\hline
\end{tabular}
\end{center}
\caption{Kernel memory loads/stores and number of floating-point operations (flop) per thread by kernel. The total number of threads is equal to the lattice volume. For kernel details see the text.}
\label{tab:memflop}
\end{table}

%%%%%%%%%%%%%%%%%%%%%%%%%%%%%%%%%%%%%%%%%%%%
%%%%%%%%%%%%%%%%%%%%%%%%%    Results   %%%%%%%%%%%%%
%%%%%%%%%%%%%%%%%%%%%%%%%%%%%%%%%%%%%%%%%%%%
\section{Results \label{resultados}}

In this section we show the benchmarks for the steepest descent Fourier accelerated code for Landau gauge
fixing in lattice QCD. The runs using the MPI implementation were performed on the Centaurus cluster at the
University of Coimbra. The Centaurus has 8 
%PS core
cores 
per node machine, each node has 24 GB of RAM, with
2 intel Xeon E5620@2.4 GHz (quad core) and has a DDR Infiniband interconnecting 
%PS the 
the various nodes.
The runs on GPU were performed at Instituto Superior T\'{e}cnico on an
NVIDIA Tesla C2070 and using version 4.1 of CUDA. The main characteristics of the
NVIDIA graphic card are summarized in Table \ref{fig:nvidia_gpu_specs}. 
Our 
%PS code GPU
GPU code 
can be downloaded from the Portuguese Lattice QCD collaboration homepage \cite{ptqcd}.

We have checked that the two independent codes produced the same configuration, within 
%PS precision machine.
machine precision.

\begin{table}[!h]
\begin{centering}
\begin{tabular}{|c|c|}
\hline
\multicolumn{2}{|c|}{\T\B NVIDIA Tesla C2070}\tabularnewline
\hline
\hline
\T\B Number of GPUs  & 1\tabularnewline
\hline
\T\B CUDA Capability & 2.0\tabularnewline
\hline
\T\B Multiprocessors (MP) & 14\tabularnewline
\hline
\T\B Cores per MP & 32\tabularnewline
\hline
\T\B Total number of cores & 448\tabularnewline
\hline
\T Global memory & 6144 MB\tabularnewline
\hline
\T\B Shared memory (per SM) & 48KB or 16KB\tabularnewline
\hline
\T\B L1 cache (per SM) & 16KB or 48KB \tabularnewline
\hline
\T\B L2 cache (chip wide) & 768KB\tabularnewline
\hline
\T\B Clock rate & 1.15 GHz\tabularnewline
\hline
\T\B  Device with ECC support & yes\tabularnewline
\hline
\end{tabular}
\par\end{centering}
\caption{NVIDIA's graphics card specifications used in this work.}
\label{fig:nvidia_gpu_specs}
\end{table}

%%%%%%%%%%%%%%%%%%%%%%%%%%%%%%%%%%%%%%%%%%%%%%%%%%%
%\subsection{Landau gauge fixing benchmarks}

\begin{figure}[H]
\begin{centering}
    \subfloat[\label{fig:time_32_3_ecc_off}]{
\begin{centering}
    \includegraphics[width=13cm]{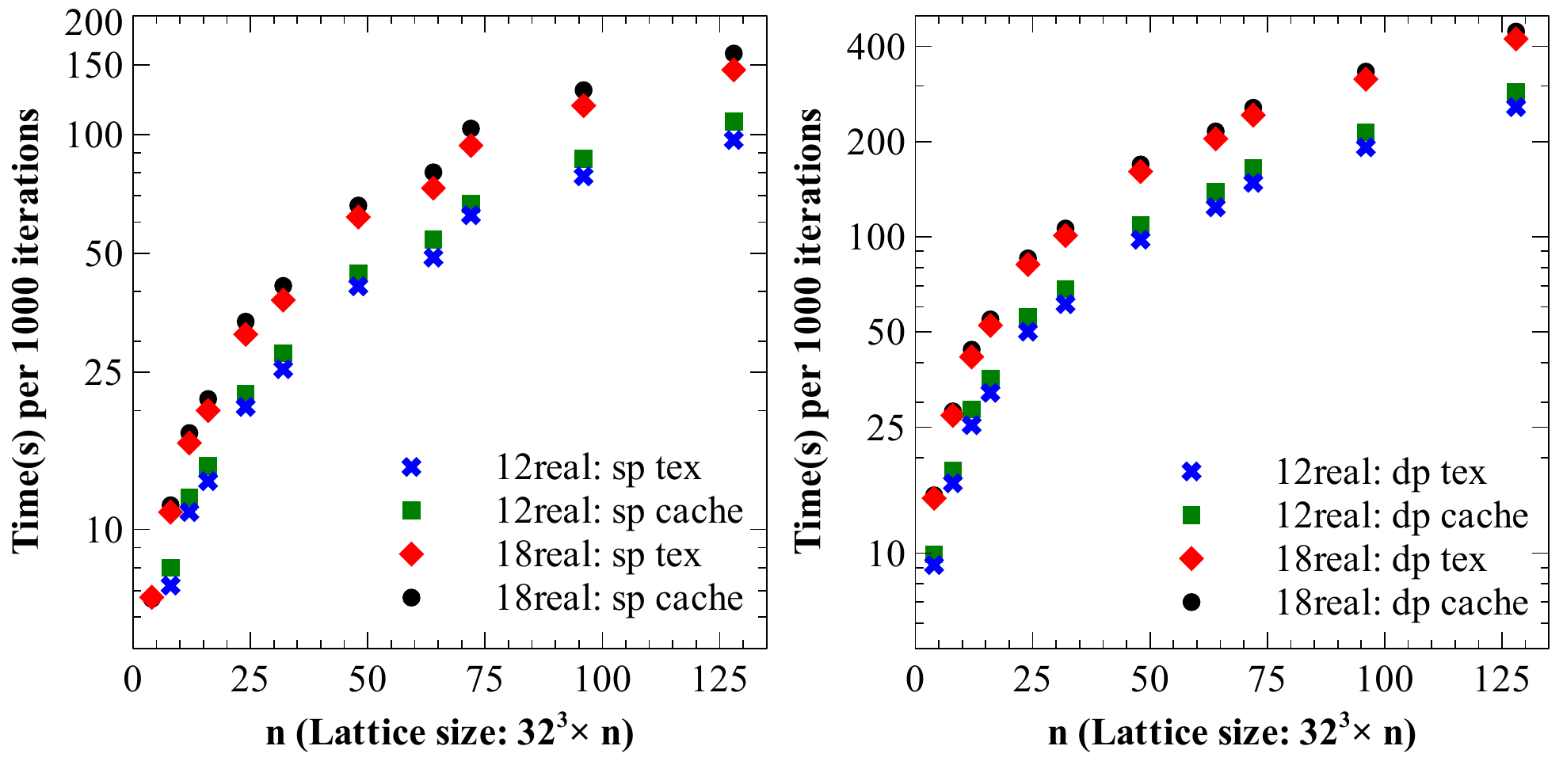}
\par\end{centering}}

    \subfloat[\label{fig:time_32_3_ecc_on}]{
\begin{centering}
    \includegraphics[width=13cm]{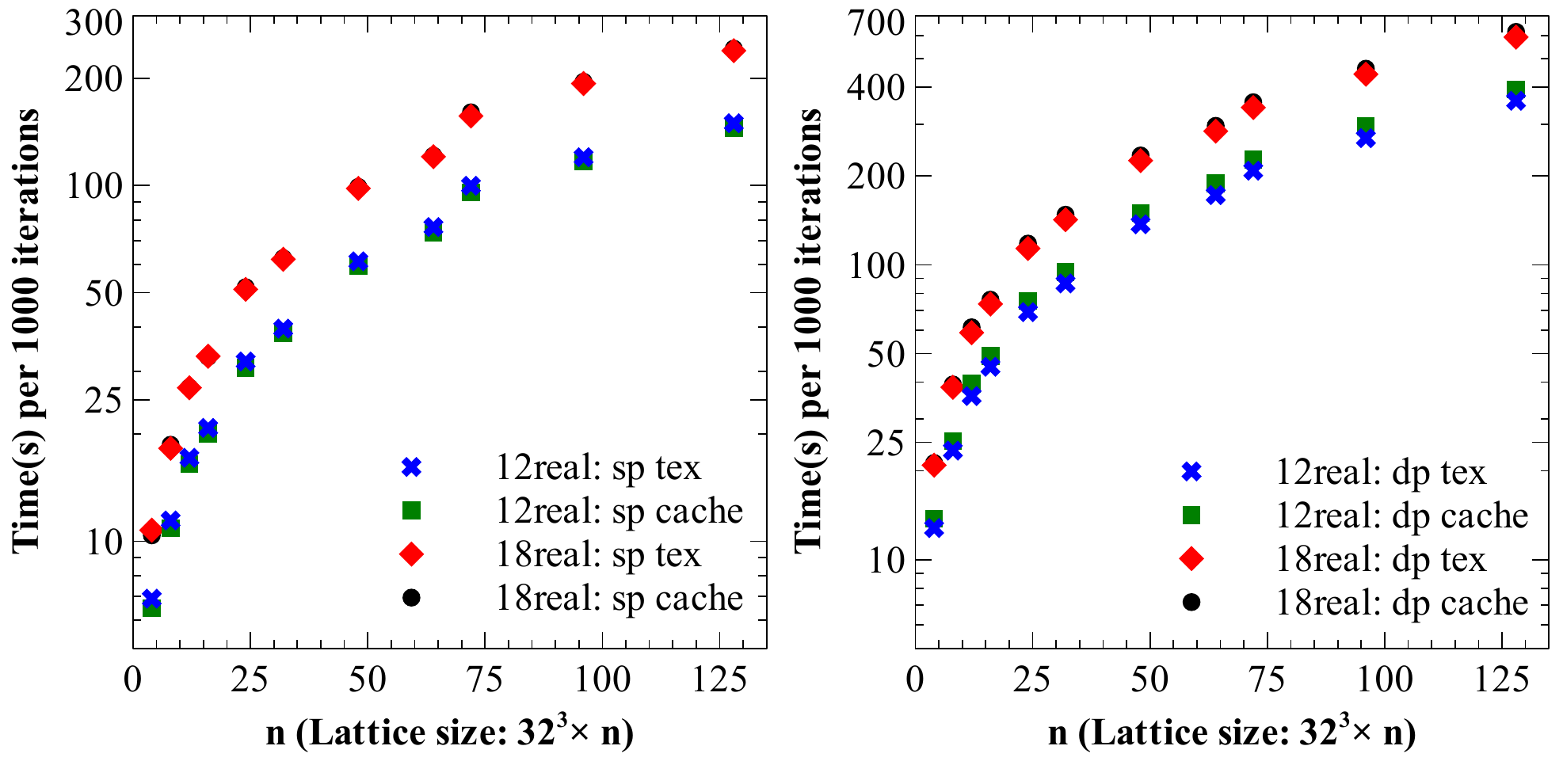}
\par\end{centering}}
\par\end{centering}
    \caption{Time (s) per 1000 iterations. \protect\subref{fig:time_32_3_ecc_off} with ECC Off and \protect\subref{fig:time_32_3_ecc_on} with ECC On. sp: single precision, dp: double precision, 18real: full SU(3) matrix, 12real: 12 real parameterization, tex: using texture memory and cache: using L1 and L2 cache memory.}
    \label{fig:time_32_3}
\end{figure}

\begin{figure}[H]
\begin{centering}
    \subfloat[\label{fig:perf_32_3_ecc_off}]{
\begin{centering}
    \includegraphics[width=13cm]{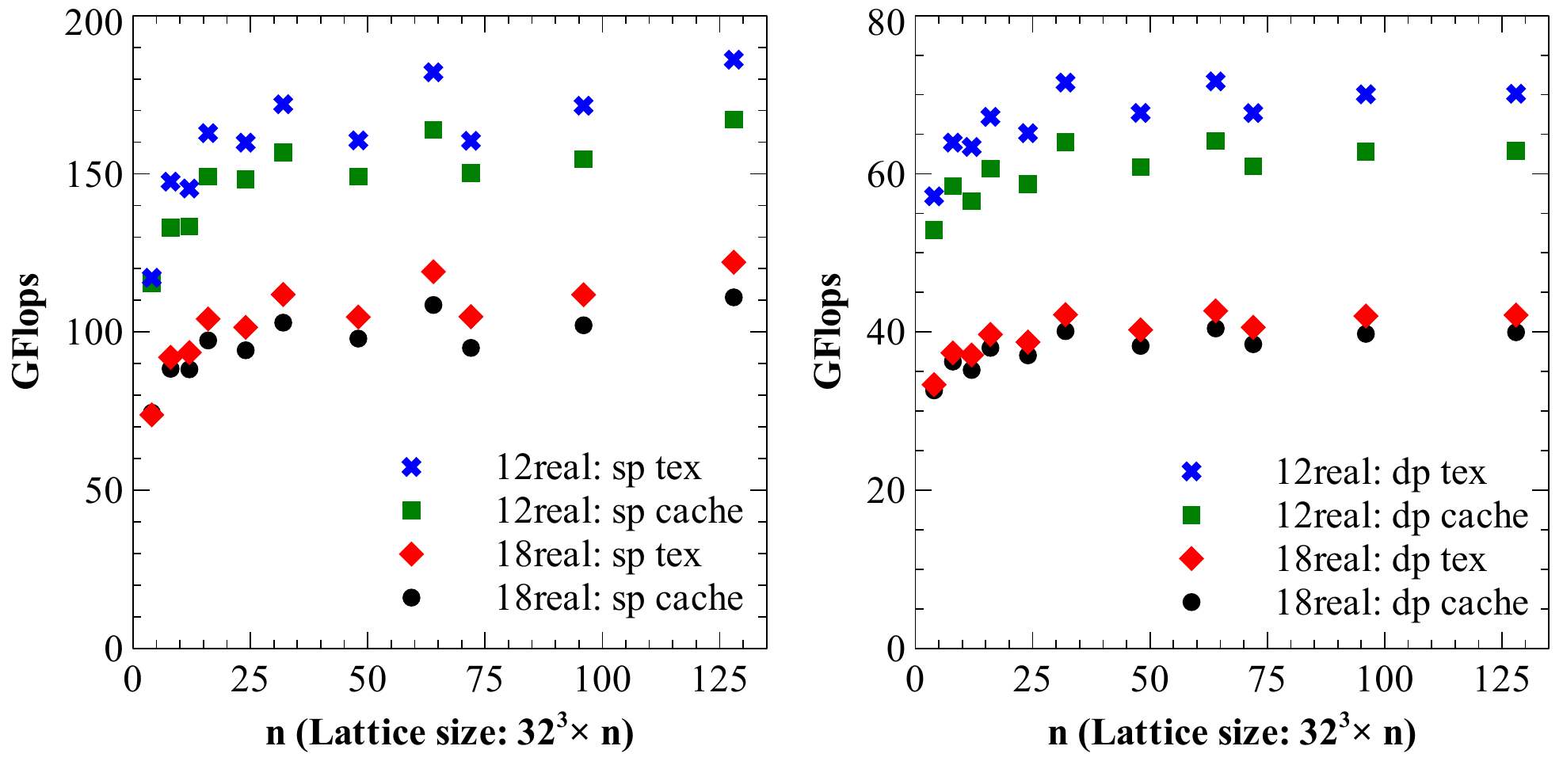}
\par\end{centering}}

    \subfloat[\label{fig:perf_32_3_ecc_on}]{
\begin{centering}
    \includegraphics[width=13cm]{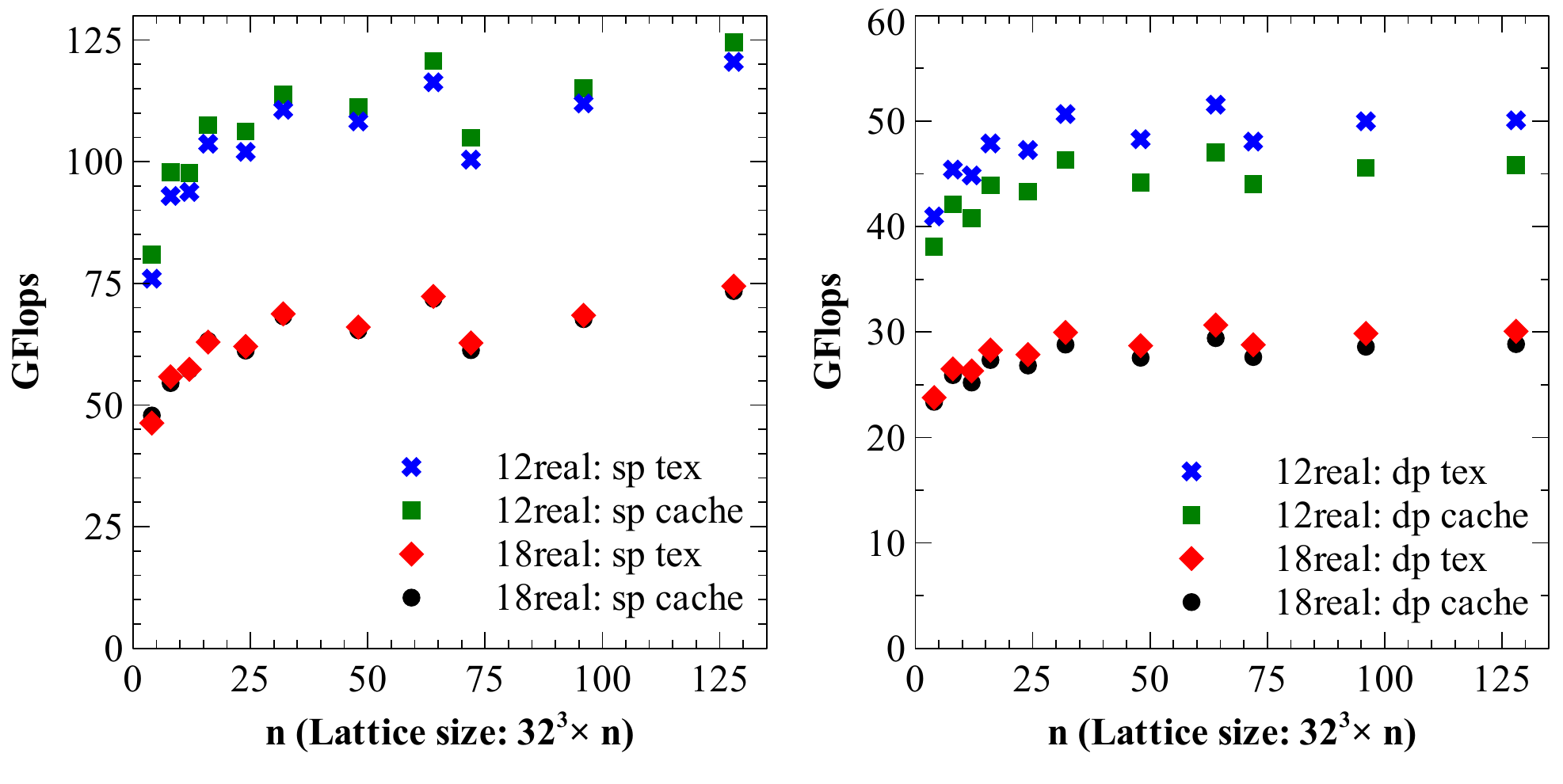}
\par\end{centering}}
\par\end{centering}
    \caption{Performance in GFlops. \protect\subref{fig:perf_32_3_ecc_off} with ECC Off and \protect\subref{fig:perf_32_3_ecc_on} with ECC On. sp: single precision, dp: double precision, 18real: full SU(3) matrix, 12real: 12 real parameterization, tex: using texture memory and cache: using L1 and L2 cache memory.}
    \label{fig:perf_32_3}
\end{figure}

For the GPU code the performance of the algorithm using a 12 parameter reconstruction and the full (18 number) representation in single and double precision was measured. Further, the GPU memory access using cache and texture memory was compared. We have also compared the performance of the three different ways to implement the 4D FFTs (1D FFTs, 3D plus 1D FFTs and 2D plus 2D FFTs) using a 12 parameter reconstruction and the full (18 number) representation in double precision, ECC off, L1 and L2 cache memory and texture memory for a $32^4$ lattice configuration with $\beta=6.0$. It turned out
that using 2D plus 2D FFTs is 39-43\% faster than using only 1D FFTs and 7-8\% faster than using 3D plus 1D FFTs. 
The best performance for our 4D problem was achieved with 2D plus 2D FFTs using cufftPlanMany(). 
In the following, we will report on the results using such a type of decomposition for the GPU simulations.

For the run tests we generated a number of $32^3 \times n$ gauge configurations at $\beta = 6.2$. In Figs. \ref{fig:time_32_3} and \ref{fig:perf_32_3} we show the outcome of the simulation
when using the GPU and compare the effects of switching from single precision to double precision. The
change in precision leads to a decrease in performance when going to double precision by a factor which can be $2.5$. The maximum of the performance code was  186 GFlops for a run
in single precision with 12 parameter reconstruction using the texture memory. It was observed that 
using texture memory instead of L1 and L2 cache memory produced always a better performance. Turning on or off
ECC  also leads to differences in performance around 1.6 - see Fig. \ref{fig:perf_32_3}.

\begin{figure}[H]
\begin{center}
    \includegraphics[width=11cm]{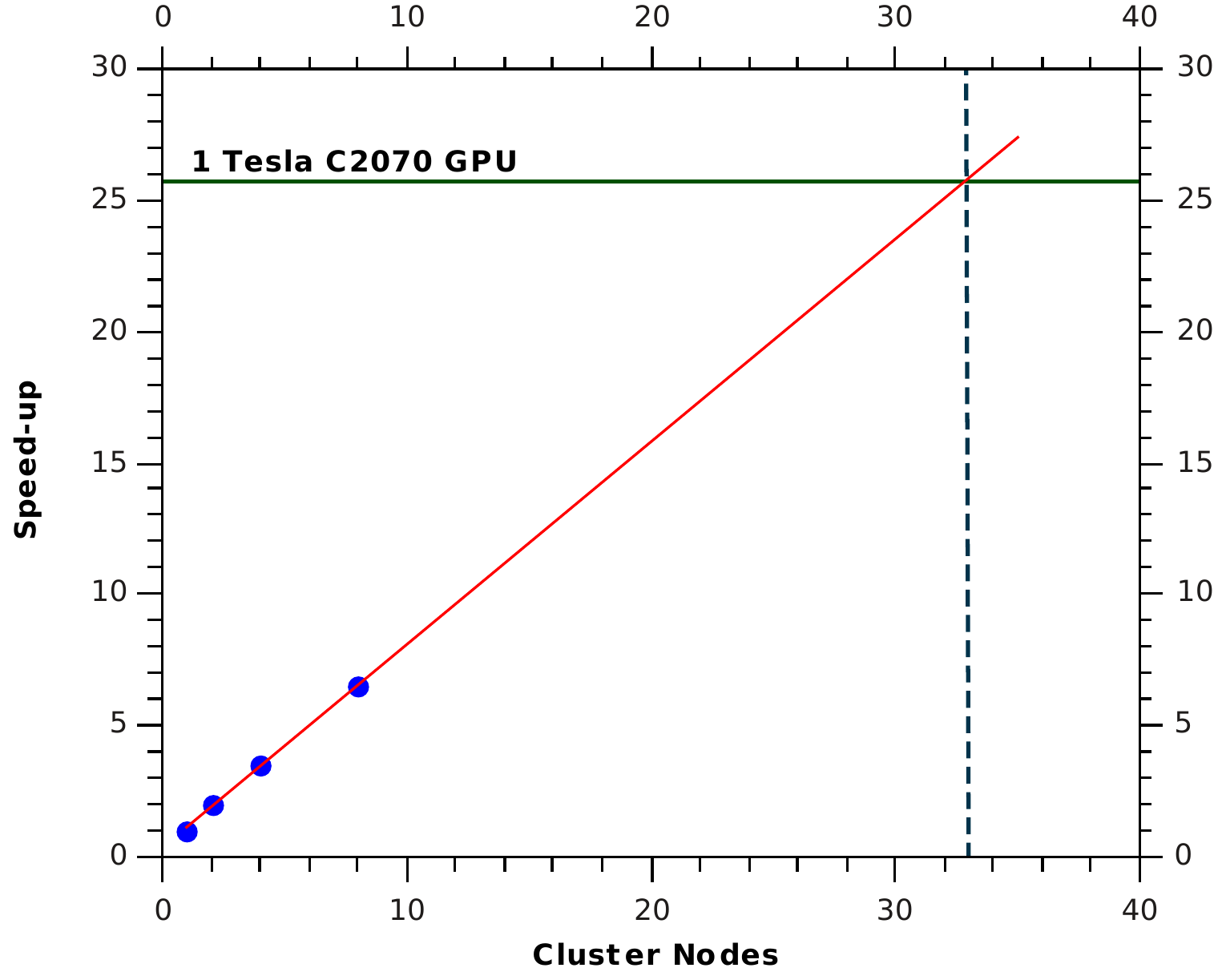}
\end{center}
\caption{Strong scaling CPU tested for a $32^4$ lattice volume and comparison with the GPU for the best 
              performance, 12 real number parameterization, ECC Off and using texture memory in double precision.
              In the Centaurus a cluster node means 8 computing cores.}
    \label{fig:cpuvsgpu}
\end{figure}

The performance of the GPU code against the parallel implementation of the same algorithm using MPI
is compared in Fig. \ref{fig:cpuvsgpu} and Table \ref{tab:CPU_GPU_time} for a $32^4$ lattice volume and for $\beta = 5.8,6.0$ and $6.2$ - see the caption of the figure and table for further details. The MPI scaling shows a good strong scaling with a linear speed-up against 
the number of computing nodes. The GPU code was clearly much faster than the MPI. The MPI implementation of the algorithm requires 32 Centaurus computing nodes, i.e. 256 CPU cores, to reproduce the performance of the GPU code.

\begin{table}[H]
\begin{center}
\begin{tabular}{|c|c|c|c|c|c|c|c|c|c|c|}
\hline
\T\B $32^4$	&		&	\multicolumn{2}{c|}{\textbf{GPU - 12 real}}	&	\multicolumn{2}{c|}{\textbf{GPU - 18 real}}	&	\multicolumn{4}{c|}{\textbf{CPU $-$ \# nodes}}		\\ \hline
\T\B $\beta$	&	iter.	&	Tex.	&	Cache	&	Tex.	&	Cache	&	8	&	4	&	2	&	1\\ \hline
\hline
\T\B 5.8	&	10427	&	637.44	&	699.49	&	1049.42	&	1086.97	&	2512.31	&	4649.83	&	8157.63	&	16323.97\\ \hline
\T\B 5.8	&	8772	&	536.31	&	588.49	&	882.80	&	914.36	&	2126.49	&	3927.31	&	6940.21	&	13766.01\\ \hline
\T\B 5.8	&	5038	&	308.02	&	338.39	&	507.03	&	525.25	&	1219.21	&	2254.74	&	3919.44	&	7835.32\\ \hline
\T\B 5.8	&	3806	&	232.67	&	255.34	&	383.02	&	396.76	&	916.74	&	1695.71	&	2962.78	&	5932.83\\ \hline
\hline
\T\B 6.0	&	3286	&	200.87	&	220.57	&	330.71	&	342.57	&	792.92	&	1466.55	&	2570.43	&	5145.40\\ \hline
\T\B 6.0	&	2954	&	180.65	&	198.17	&	297.29	&	307.93	&	719.52	&	1318.45	&	2309.70	&	4615.02\\ \hline
\T\B 6.0	&	2320	&	141.84	&	155.73	&	233.50	&	241.86	&	561.31	&	1034.58	&	1823.44	&	3638.41\\ \hline
\T\B 6.0	&	4189	&	256.31	&	281.04	&	421.65	&	436.64	&	1014.53	&	1877.51	&	3276.73	&	6582.08\\ \hline
\hline
\T\B 6.2	&	1083	&	66.42	&	72.69	&	109.00	&	112.92	&	263.78	&	484.91	&	849.60	&	1690.28\\ \hline
\T\B 6.2	&	1342	&	82.06	&	90.07	&	135.07	&	139.92	&	325.81	&	600.56	&	1046.26	&	2087.11\\ \hline
\T\B 6.2	&	5403	&	330.33	&	362.54	&	543.83	&	563.19	&	1307.71	&	2403.98	&	4201.31	&	8462.38\\ \hline
\T\B 6.2	&	765	&	46.78	&	51.35	&	77.01	&	79.77	&	185.05	&	342.56	&	599.99	&	1272.64\\ \hline
\hline
\end{tabular}
\end{center}
\caption{Number of iterations and time (s) to perform Landau gauge fixing for a lattice volume of $32^4$ in double precision. This was done for three values of $\beta$ with four 
%PS sets of 
configurations each. The GPU code was tested in a Tesla C2070 GPU with ECC Off with and without the 12 parameter reconstruction for SU(3), the CPU code was tested in the Centaurus cluster, 
%PS node, each one with 
where each node has
8 computing cores and DDR Infiniband interconnecting the 
%PS the 
various nodes, using 
%PS full SU(3) matrix.
a full SU(3) matrix representation.
The value of $\alpha$ is $0.08$ and the stop criterion is $\theta< 1\times10^{-15}$.}
\label{tab:CPU_GPU_time}
\end{table}

%%%%%%%%%%%%%%%%%%%%%%%%%%%%%%%%%%%%%%%%%%
%%%%%%%%%%%%%%%%%%%%%%%%%    Conclusion   %%%%%%%%%
%%%%%%%%%%%%%%%%%%%%%%%%%%%%%%%%%%%%%%%%%%
\section{Conclusion \label{conclusoes}}

We have compared the performance of the GPU implementation of the Fourier accelerated steepest descent
Landau gauge fixing algorithm using CUDA with a standard MPI implementation built on the Chroma library.
The run tests were done on $32^4$ and $\beta = 5.8,6.0$ and $6.2$ pure gauge configurations, generated
using the standard Wilson action.

In order to optimize the GPU code, its performance was investigated on $32^3 \times n$ gauge configurations.
The runs on a C2070 Tesla show that, for a 4D dimension lattice, the best performance was achieved 
with 2D plus 2D FFTs using \lstinline!cufftPlanMany()! and using a 12 real parameter reconstruction with
texture memory. From all the runs using a C2070 Tesla GPU, peak performance was measured
as 186/71 GFlops for single/double precision. From the performance point of view, a run on a single GPU delivers
the same performance as the 
%PS MPI
CPU code when running on 32 nodes (256 cores),
if one assumes a linear speed-up behavior.

%As for the generation of gauge configurations, the use of GPUs reduces considerably the time of a simulation.
%Presently, the main limitation of the GPUs both for gauge generation and for Landau gauge fixing is its limited
%memory size. For the Tesla C2070, the memory is 6GB.  
%PS
%This allows to consider lattice volumes up to
%NC
%$56^4$ using 12 real number parameterization in double precision to perform gauge fixing
%on a single GPU. Of course, this is not a competitive lattice volume in present days. In order to consider the large lattice volumes being worked out in present days (like $96^4$, see, for example, \cite{Bogolubsky:2009dc}) on GPU architectures, one has to consider a multi-GPU algorithm. We leave this for a future work.

As for the generation of gauge configurations, the use of GPUs reduces considerably the time of a
simulation. At present, the main limitation of the GPUs both for gauge generation and for Landau gauge
fixing is its limited memory size. For the Tesla C2070, the memory is 6GB. This allows us to consider lattice
volumes up to $56^4$ using 12 real number parameterization in double precision to perform gauge fixing on a
single GPU.  In order to consider larger lattice volumes on GPU architectures,
one has to consider a multi-GPU implementation. We leave this for future work.

%%%%%%%%%%%%%%%%%%%%%%%%%%%%%%%%%%%%%%%%%%%%
%%%%%%%%%%%%%%%%%%%%%%%%%%% acknowledgments %%%%%%%
%%%%%%%%%%%%%%%%%%%%%%%%%%%%%%%%%%%%%%%%%%%%
%\acknowledgments
\section*{Acknowledgments}
We thank B\'{a}lint Jo\'{o} and Michael Pippig for discussions about Chroma and PFFT libraries respectively. In particular, we thank Michael Pippig for extending his library for 4-dimensional FFTs. 

This work was partly funded by the FCT contracts,  POCI/FP/81933/2007, 
CERN/FP/83582/2008, PTDC/FIS/100968/2008, CERN/FP/109327/2009, CERN/FP/116383/2010 and CERN/FP/123612/2011.

%PS Nuno Cardoso is also supported by FCT under the contract SFRH/BD/44416/2008.
Nuno Cardoso and Paulo Silva are supported by the FCT under contracts SFRH/BD/44416/2008 and SFRH/BPD/40998/2007 respectively.
We would like to thank NVIDIA Corporation for the hardware donation used in this work via the Academic Partnership 
program.

%================================================================
%================================================================
\bibliographystyle{elsarticle-num}

\bibliography{bib}

\end{document}